\documentclass[%
 aip,
 amsmath,amssymb,
reprint,%
]{revtex4-1}

\usepackage[T1]{fontenc}
\usepackage[english]{babel}
\usepackage{graphicx}
\usepackage{epsfig}
\usepackage{dcolumn}
\usepackage{bm}
\usepackage{epstopdf}
\usepackage{amsmath}
\usepackage{amssymb}
\usepackage{times}
\usepackage{setspace}
\usepackage{verbatim}
\usepackage{color}
\usepackage{hyperref}
\usepackage{mathrsfs}
\usepackage{mathptmx}

\begin{document}
\title{Dynamics of a driven confined polyelectrolyte solution}
\author{Debarshee Bagchi}
\email[E-mail address:]{debarshee.bagchi@northwestern.edu}
\affiliation{Department of Materials Science and Engineering, Northwestern University, Evanston, IL 60208, United States}
\author{Monica Olvera de la Cruz}
\email[E-mail address: ]{m-olvera@northwestern.edu}
\affiliation{Department of Materials Science and Engineering, Northwestern University, Evanston, IL 60208, United States}
\affiliation{Department of Physics and Astronomy, Northwestern University, Evanston, IL 60208, United States}
\date{\today}

\begin{abstract}
The transport of polyelectrolytes confined by oppositely charged surfaces and driven by a constant electric field is of interest in studies of DNA separation according to size. 
Using molecular dynamics simulations that include surface polarization effect, we find that the mobilities of the polyelectrolytes and their counterions change non-monotonically 
with the confinement surface charge density. For an optimum value of the confinement charge density, efficient separation of polyelectrolytes can be achieved over a wide range of 
polyelectrolyte charge due to the differential friction imparted by the oppositely charged confinement on the polyelectrolyte chains. Furthermore, by altering the placement of the 
charged confinement counterions, enhanced polyelectrolyte separation can be achieved by utilizing surface polarization effect due to dielectric mismatch between the media inside 
and outside the confinement.
\end{abstract}
\maketitle


\section{Introduction}
Electrophoretic techniques for separating and sequencing DNA, proteins, other biomolecules and synthetic polymers according to size or charge, have been extensively used in recent 
times for genome analysis \cite{dovichi2000capillary, kounovsky2017electrostatic}, clinical diagnostics \cite{wuethrich2019decade}, forensic investigations 
\cite{anastos2005capillary}, industrial processes \cite{dulffer1990capillary} and many other applications. After it was realized that the mobility of DNA in free solution is 
independent of its length \cite{olivera1964electrophoresis,stellwagen1997free}, besides the conventional gel electrophoresis technique \cite{testa1965separation}, which is 
insensitive to separate long molecular weight polyelectrolytes \cite{de1986electrophoresis,shaffer1989dynamics}, several other separation techniques have been subsequently 
developed, such as the capillary electrophoresis \cite{barron1994transient,hubert1996theory}, pulsed-field gel electrophoresis \cite{de1990dynamics}, pulsed-field gradient
gel electrophoresis \cite{schwartz1984separation}, pulsed-field capillary gel electrophoresis \cite{sudor1994separation}, microchip capillary electrophoresis 
\cite{liu2003microfabricated}, dielectrophoresis \cite{chou2002electrodeless}, entropic traps \cite{han2000separation}, nanopatterned surface \cite{seo2004dna} and flat surface 
electrophoresis \cite{pernodet2000dna}, to name a few. Each of these techniques has their advantages and shortcomings, based on factors such as separation resolution, speed, 
reproducibility, cost effectiveness, sample requirements, automation, ease of integration etc. For example, pulsed-field gel electrophoresis, the most widely used method for large 
DNA molecules, suffers from problems of slow speed and difficulty of automation. On the other hand, conventional capillary electrophoresis has the advantages of being fast and 
efficient with low sample requirements and is easily integrable, but on the downside, separation sizes are limited to the kbp range for DNA electrophoresis, along with the 
difficulty of loading a high viscosity polymer solution inside a narrow capillary. The problem of loading a high viscosity polymer solution inside a narrow capillary 
can be circumvented by using an ultradilute polymer solution, as in the transient entanglement coupling mechanism \cite{barron1994transient}, but then the resolution for short DNA 
fragments becomes poor. Moreover, for these methods, the separation resolution is restricted by the properties of the sieving matrix. Recently, methods without sieving matrices 
have been proposed, such as by using nanopattered structures, but these techniques are still in their infancy and have low separation speed and resolution. As such, newer and more 
efficient techniques for separating DNA and other bio-macromolecules are being continuously pursued (for reviews on theoretical, computational, and experimental works on 
conventional and more recent electrophoretic techniques, see \cite{barron1995dna,viovy2000electrophoresis,heller2001principles, 
lin2005nanomaterials,baldessari2006electrophoresis,kleparnik2007dna,dorfman2010dna,viefhues2017dna}).

In this article, we show that a separation technique without involving sieving matrices or pulsed electric fields, and readily integrable with nanofluidic systems, can be 
conceptualized using a charged confinement. To optimize the resolution of the separation technique, we study the dynamics of negatively charged polymer solutions under confinement 
by a positively charged cylindrical nanochannel, and driven by a constant external electric field. We also examine the effect of surface polarization due to mismatch of 
dielectric constants between the media inside and outside the charged confinement. Previous studies have shown that polarization effect often leads to quantitative as well as 
qualitative changes in the physical properties of charged systems \cite{nguyen2019manipulation, bagchi2020surface}. Here, we demonstrate when and why the properties of the 
confined polyelectrolyte solution are affected by dielectric mismatch, and use the results of our analysis to optimize the separation technique. It is clear that the molecular 
weight dependent friction of the negatively charged polyelectrolytes, adsorbed on to an oppositely charged surface, plays a key role in separating the polyelectrolyte chains, a 
critical step for reading the genome of an organism.

Motivated by the generation of strong electrostatic correlations when charges are present in a low dielectric environment, such as when charged particles dissolve in organic 
solvents in contact with aqueous solution \cite{leunissen2007ion}, or when polyelectrolytes are confined in cavities with ions in the low dielectric media 
\cite{nguyen2019manipulation}, we consider two models, referred to as Model-I and Model-II. In Model-I, the counterions of the positively charged surface are in the high 
dielectric (aqueous) medium and in Model-II, the counterions are in the low dielectric medium (this happens when, say, intercalated ions in graphene are in contact with an aqueous 
solution or when the charged particles in oil release their protons at the interface with water (see Fig. 2 in \cite{leunissen2007ion}).

Using coarse-grained molecular simulations, we demonstrate that the confined polyelectrolyte solution has rich transport properties that can be conveniently controlled by 
tuning the surface charge density (SCD) on the confinement in both Model-I and Model-II. In particular, when polarization effect is taken into account in Model-II, polyelectrolytes 
can be separated with high resolution due to image charge effects and strong ionic correlations, if the SCD is of the opposite sign to that of the polyelectrolyte charge. Since the 
key factor for separation is the differential friction experienced by polyelectrolyte chains of different charges (or equivalently, different sizes, as in our case) due to the 
charged confinement, we concentrate on the case for which the confinement surface and the polyelectrolyte are oppositely charged (although for completeness, in the SI we present 
data of the mobility when the confinement surface charge is of the same sign as that of the polyelectrolyte). Besides electrophoresis, our study is also relevant for understanding 
confinement effects on the dynamics of charged soft-matter systems and in the design of smart nanofluidic devices \cite{napoli2010nanofluidic}.

\section{The confined polyelectrolyte model} 
For our course-grained molecular dynamics simulation, the polyelectrolyte is modeled as a linear bead-spring chain with explicit counterions in an implicit water-like solvent. This 
polyelectrolyte solution is confined inside an impenetrable cylindrical confinement that is comprised of discrete Lennard-Jones (LJ) beads. Each confinement surface bead carry a 
partial charge and the confinement has a charge density denoted by $\Sigma$. All the beads on each polyelectrolyte chain are negatively charged with one unit of elementary charge
$-e$. There are typically $N_c = 10$ polyelectrolyte chains inside the confinement, each with $N_m = 40$ monomers, for most of our simulations. The consecutive beads in the
polyelectrolyte chain are connected by finite extensible nonlinear elastic (FENE) bonds. For overall electroneutrality, there are $N_{cions} = N_c N_m$ positively charged 
counterion beads inside the confinement. The cylindrical confinement has a length $L=L_z$ and radius $R$, with periodic boundary conditions in the z-direction. The positive charge 
on the cylinder is balanced by an equivalent number of monovalent negatively charged beads, that are placed inside the confinement for Model-I, and outside the confinement for 
Model-II. All the charged species in the simulation interact via long-ranged Coulomb interactions, and we set the Bjerrum length $l_B = 0.7 ~nm \approx 2.33 \sigma$ for water with 
uniform dielectric constant $ \epsilon_r = 80$; $\sigma$ is the length unit of the coarse-grained system. We apply a constant electric field along the axis of the cylinder $\vec E 
\equiv (0,0,E_z)$, with $E_z = 0.025$ (in reduced units), and corresponds to approximately $3 \times 10^7$ V/m. Electric fields of similar magnitude are typically used in 
coarse-grained simulation studies of DNA electrophoresis \cite{pernodet2000dna}. For our simulations, we have used the Verlet-velocity scheme in the canonical ensemble with 
Langevin thermostat at reduced temperature $T = 1$ and friction coefficient $\gamma = 1$. The simulations are performed using the software package Large-scale Atomic/Molecular 
Massively Parallel Simulator (LAMMPS) \cite{plimpton1995fast}, and for computing the induced charges due to dielectric mismatch, we employ an efficient algorithm known as the 
Induced Charge Computation (ICC) using the boundary element method (see \cite{nguyen2019incorporating} and references therein for more details).
\begin{figure}[htb]
\centering
{\includegraphics[width=6.25cm]{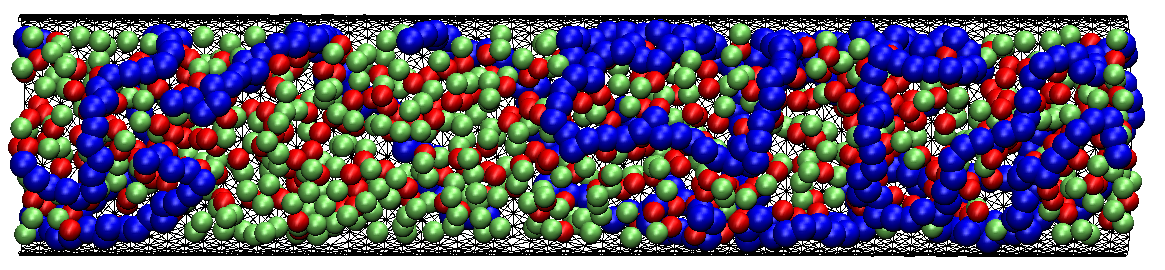}}
\caption{Model-I: Typical configuration of polyelectrolyte chains (blue), counterions (red), and surface counterions (green) inside a cylindrical confinement.}
\label{fig:model-I}
\end{figure}

To characterize the dynamics of the confined polyelectrolyte solution, the main quantity of interest is the mobility $\mu = \dfrac{v_z}{q E_z}$, where $v_z$ is the 
average velocity of the charged species parallel to the direction of the applied external field. We denote the mobility of the polyelectrolyte chains and their counterions as 
$\mu_P$ and $\mu_C$ respectively. The mobility of the polyelectrolyte chains depends on the complex interplay of (a) the applied electric field $\vec E$, (b) the charge 
on each chain $-N_m e$, (c) the charge density on the confinement $\Sigma$, and (d) the counterion screening (cs), which we define as the fraction $f_{cs}$ of the total 
counterions that are present within a distance of $1.5\sigma$ from any monomer of a polyelectrolyte chain. The counterion screening fraction $f_{cs}$ is an important quantity for 
understanding the transport features of the system since it is related to the effective charge of the polyelectrolyte chains.

\section{Simulation Results for Model-I}

\subsection{Mobilities without Polarization effect}
\begin{figure}[htb]
\hskip-0.13cm
{\includegraphics[width=4.50cm]{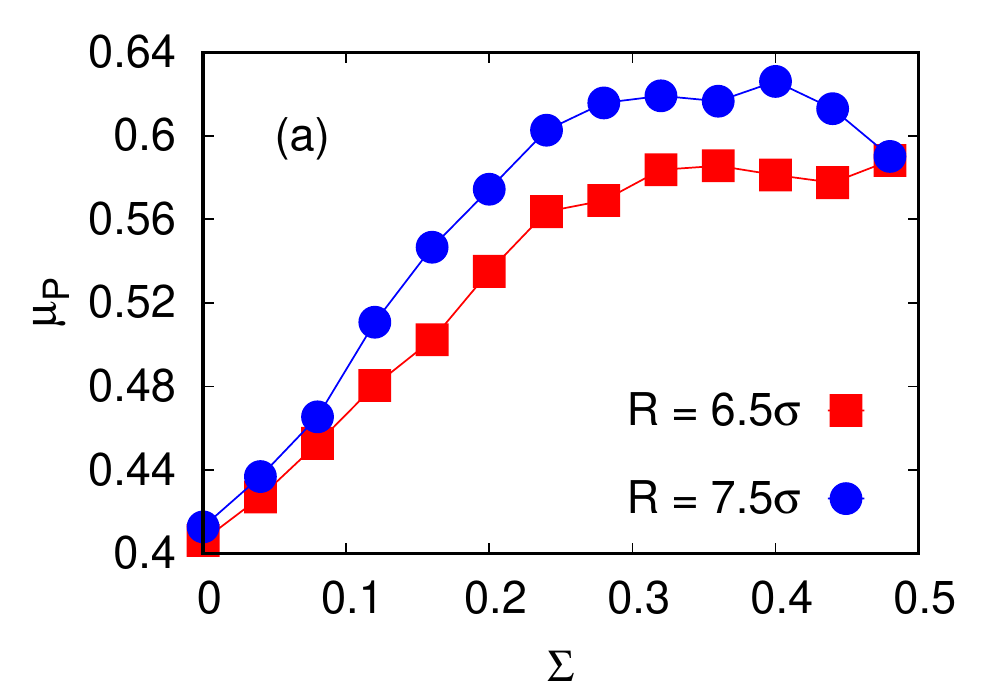}}\hskip-0.23cm
{\includegraphics[width=4.50cm]{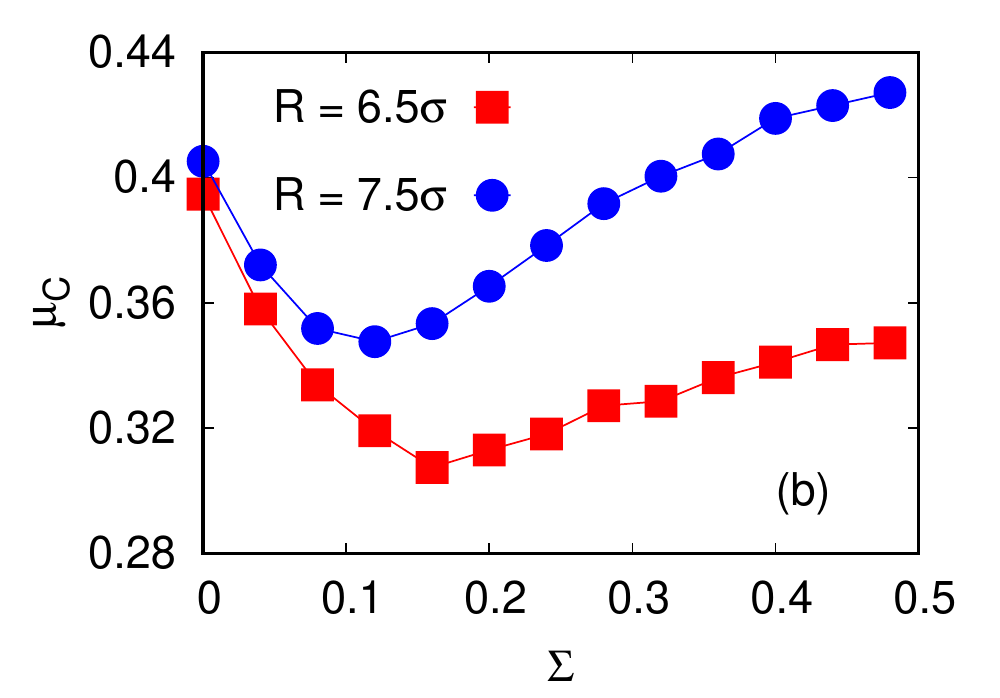}}\\
{\includegraphics[width=4.40cm]{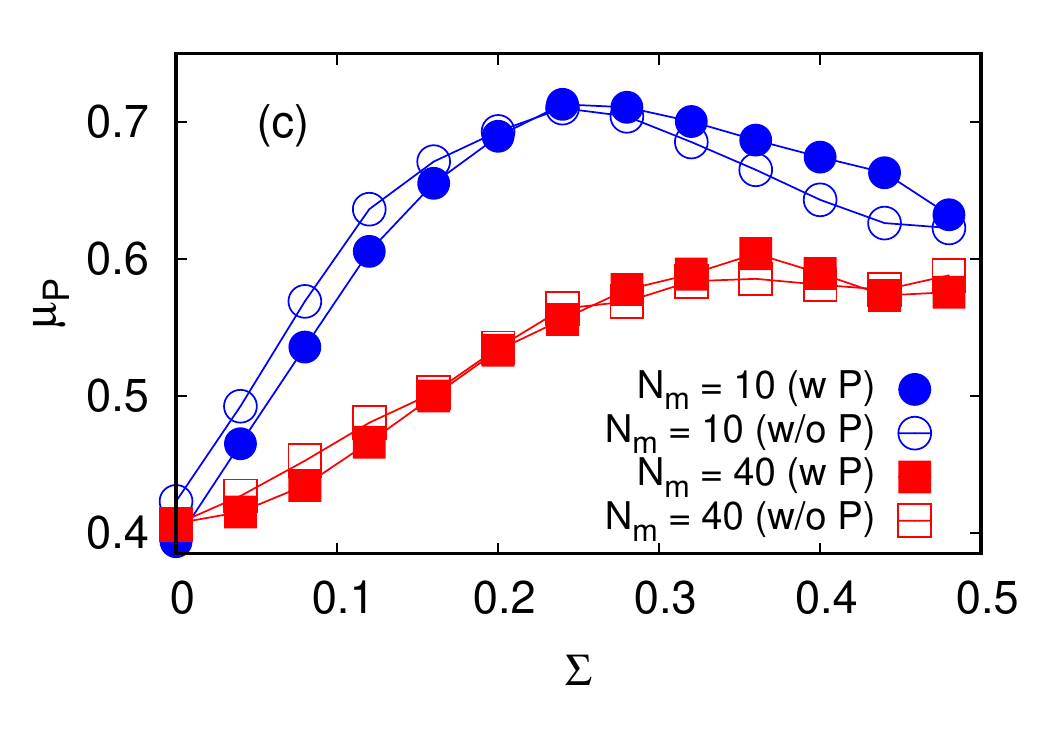}}\hskip-0.2cm
{\includegraphics[width=4.40cm]{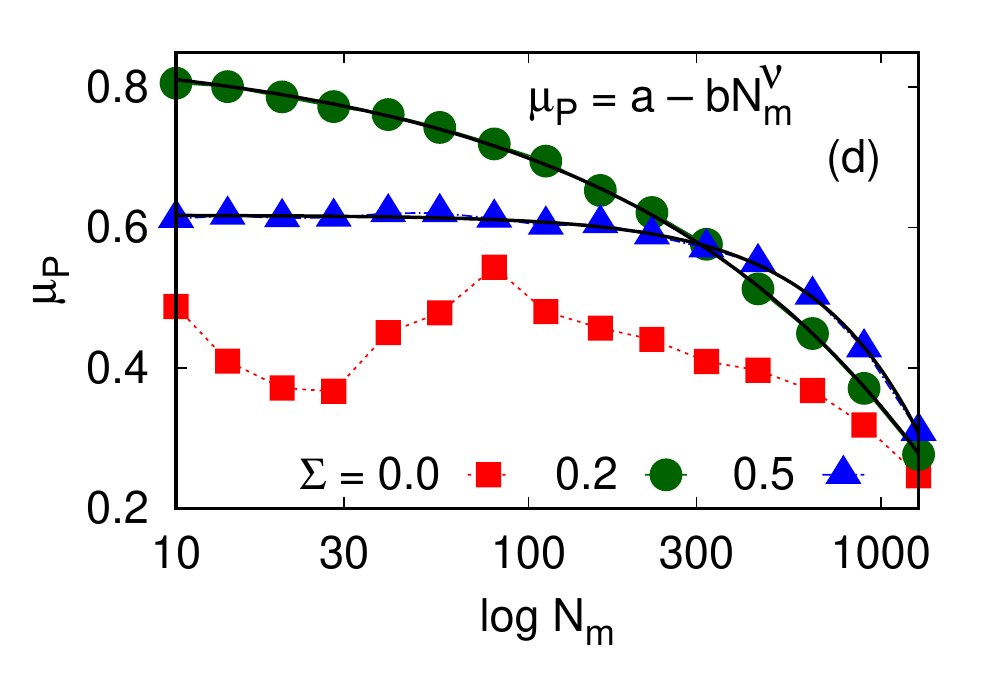}}
\caption{(a) Polyelectrolyte mobility $\mu_P$ and (b) counterion mobility $\mu_C$ as a function of surface charge density $\Sigma$, for two different confinement radii, $R = 
6.5\sigma$ and $R = 7.5\sigma$. (c) Polyelectrolyte mobility $\mu_P$ for two polyelectrolyte solutions, with $N_m = 10$ and $40$ charged monomers on each polyelectrolyte chain, 
without (w/o P) and with polarization (w P) effect. Total number of chains used $N_c = 10$. (d) Variation of polyelectrolyte single chain mobility $\mu_P$ with changing $N_m$, for 
confinement charge density $\Sigma = 0, 0.2$, and $0.5$ $C/m^2$. The continuous lines for $\Sigma > 0$ represent the function $a-bN_m^{\nu}$.}
\label{fig:poschconf}
\end{figure}

We begin our analysis with Model-I, for which we first benchmark the dynamics of the confined polyelectrolyte solution without polarization effect. In Fig. \ref{fig:poschconf}a 
and b, the average mobilities of the polyelectrolyte chains and their counterions are shown, as the SCD $\Sigma$ on the confinement is varied, for confinement radii $R = 
6.5\sigma$ and $7.5\sigma$. The value of the electric field $\vec E$ is chosen to be in the linear regime, such that the mobilities $\mu_P$ and $\mu_C$ are practically independent 
of the $\vec E$ field \textcolor{blue}{(Fig. S1, SI Appendix)}. From Fig. \ref{fig:poschconf}, the mobilities of the polyelectrolyte chains $\mu_P$ and their counterions $\mu_C$ 
are found to change non-monotonically with $\Sigma$. 

The non-monotonic change of the polyelectrolyte mobility can be understood as a combined effect of counterion release and adsorption of the polyelectrolyte chains on to the surface 
of the confinement. Initially, as the SCD $\Sigma$ is increased from zero, the number of counterions screening the polyelectrolyte chains decreases \textcolor{blue}{(Fig. S2, SI 
Appendix)}, and consequently, the effective charge on the chains increases. This leads to an increase in the polyelectrolyte mobility $\mu_P$. However, when the SCD becomes very 
high, the negatively charged polyelectrolyte chains get adsorbed on the positively charged surface of the confinement, and hence $\mu_P$ decreases for large $\Sigma$. For the 
confinement of smaller radius, $R = 6.5 \sigma$, the polyelectrolyte mobility $\mu_P$ is lower due to the fact that the fraction $f_{cs}$ is higher as compared to $R = 7.5 \sigma$ 
\textcolor{blue}{(Fig. S2, SI Appendix)}. 

The non-monotonicity of the counterion mobility can be rationalized by noting that, as SCD $\Sigma$ increases, the effective charge of the polyelectrolyte chains also increases 
(since $f_{cs}$ decreases, \textcolor{blue}{Fig. S2, SI Appendix}), and as the chains move through the confinement driven by the $\vec E$ field, they drag a part of the 
counterions with them. This drag force on the counterions due to the polyelectrolyte chains decreases the average counterion mobility $\mu_C$ initially. However, as $\Sigma$ 
increases further, the polyelectrolyte chains release the counterions even more ($f_{cs}$ decreases further). As such, the counterions are not dragged by the polyelectrolyte chains 
anymore and this leads to the increase of the counterion mobility $\mu_C$. We have further verified that such non-monotonic behavior of the counterions is also observed in 
multivalent size-symmetric electrolyte solutions, but not in monovalent electrolytes \textcolor{blue}{(Fig. S3, SI Appendix)}, and this can be explained following similar 
arguments.

One can also explain why the $\Sigma$ value at which the curves for the counterion and the polyelectrolyte mobilities show non-monotonic behavior, is lower for the cylinder with 
the larger radius $R = 7.5\sigma$. For a larger confinement radius, the counterion screening fraction $f_{cs}$ is lower \textcolor{blue}{(Fig. S2, SI Appendix)}, and therefore, 
the chains get absorbed on to the confinement at a smaller $\Sigma$, and $\mu_P$ decreases. For the same reason, the counterion mobility $\mu_C$, for $R=7.5\sigma$, starts 
to increase at a lower $\Sigma$, since the drag due to the polyelectrolyte chains is lower in a larger confinement. We have checked that similar non-monotonic mobility trends are 
also observed for confinements with different lengths, keeping the same radius $R=6.5\sigma$ \textcolor{blue}{(Fig S4, SI Appendix)}.

Note that, the mobilities of the polyelectrolyte and the counterions, $\mu_P$ and $\mu_C$, are equal for an uncharged confinement, but for a charged confinement in the range $0 < 
\Sigma < 0.5 ~C/m^2$, we have $\mu_P > \mu_C$ (as in Fig. \ref{fig:poschconf}a,b). Thus, the symmetric flow of the negative polyelectrolyte chains and the positive counterions 
is broken by the charged confinement. 

\begin{figure*}[htb]
\centering
{\includegraphics[width=4.8cm]{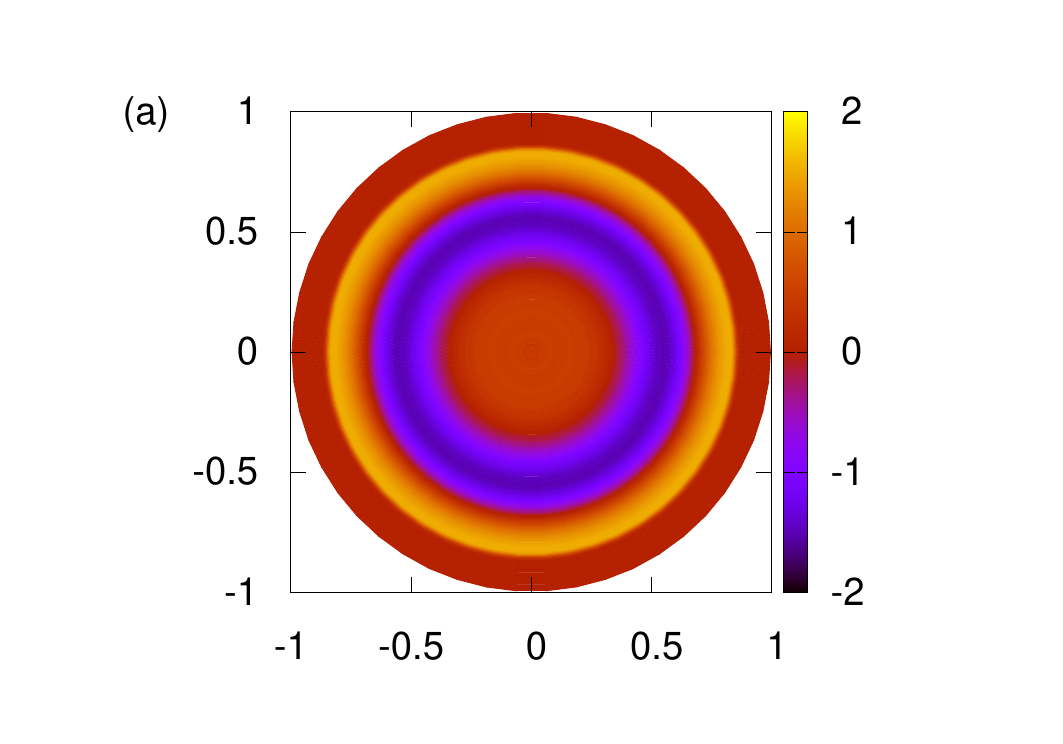}}
{\includegraphics[width=4.8cm]{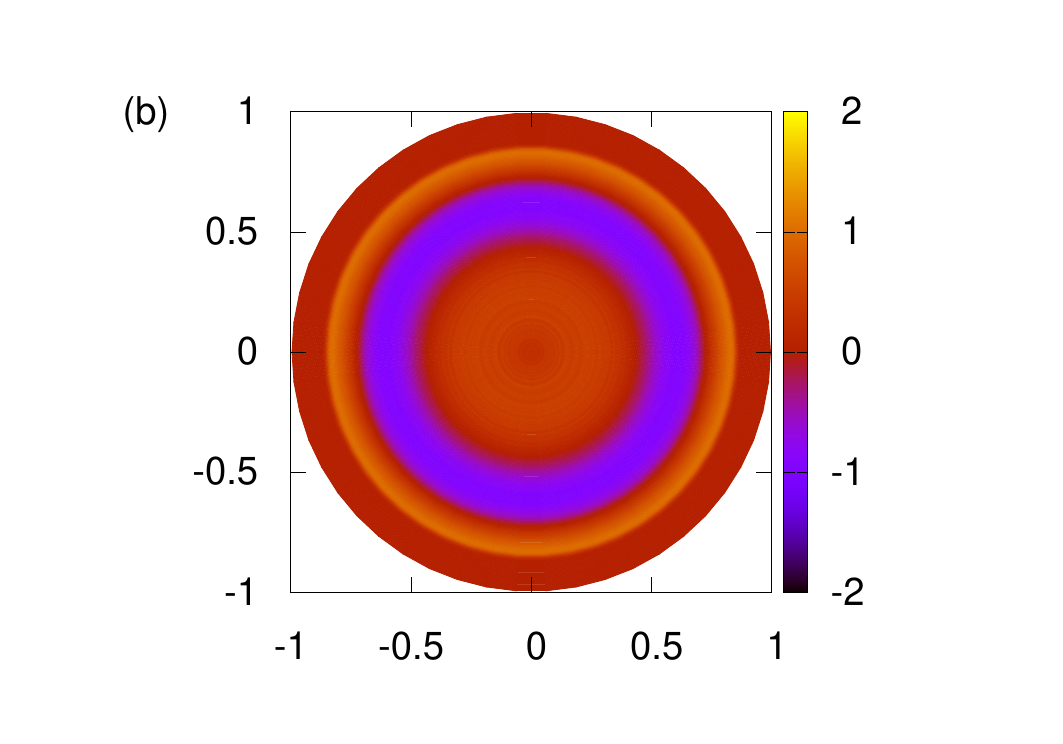}}
{\includegraphics[width=4.8cm]{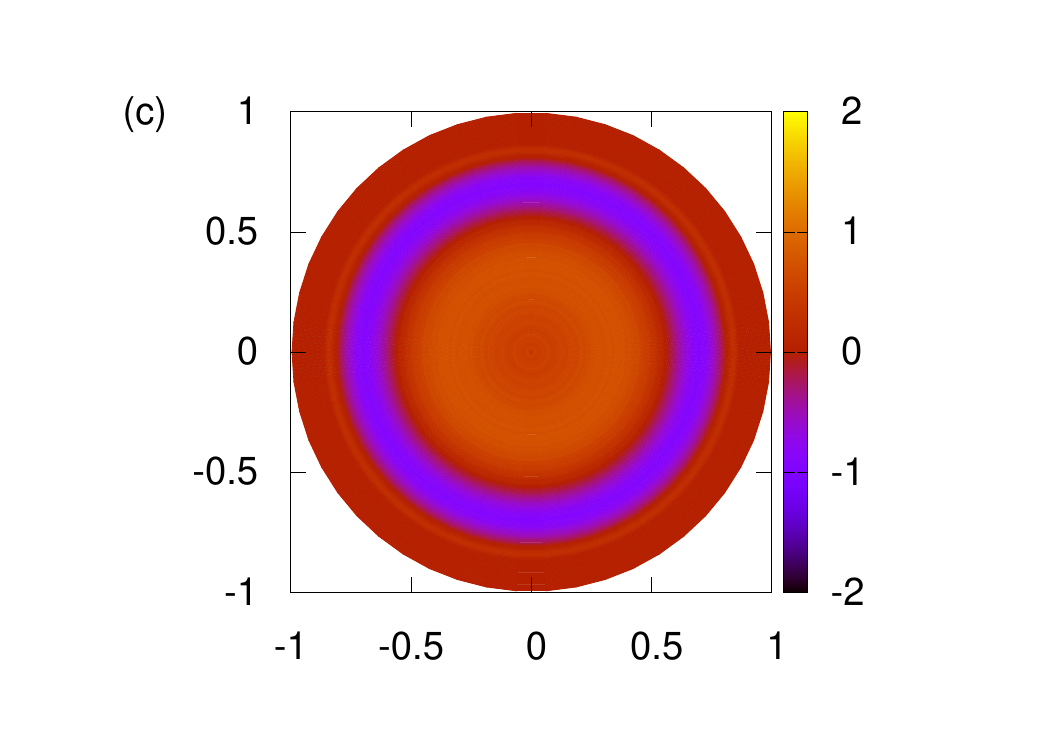}}\\
{\includegraphics[width=4.8cm]{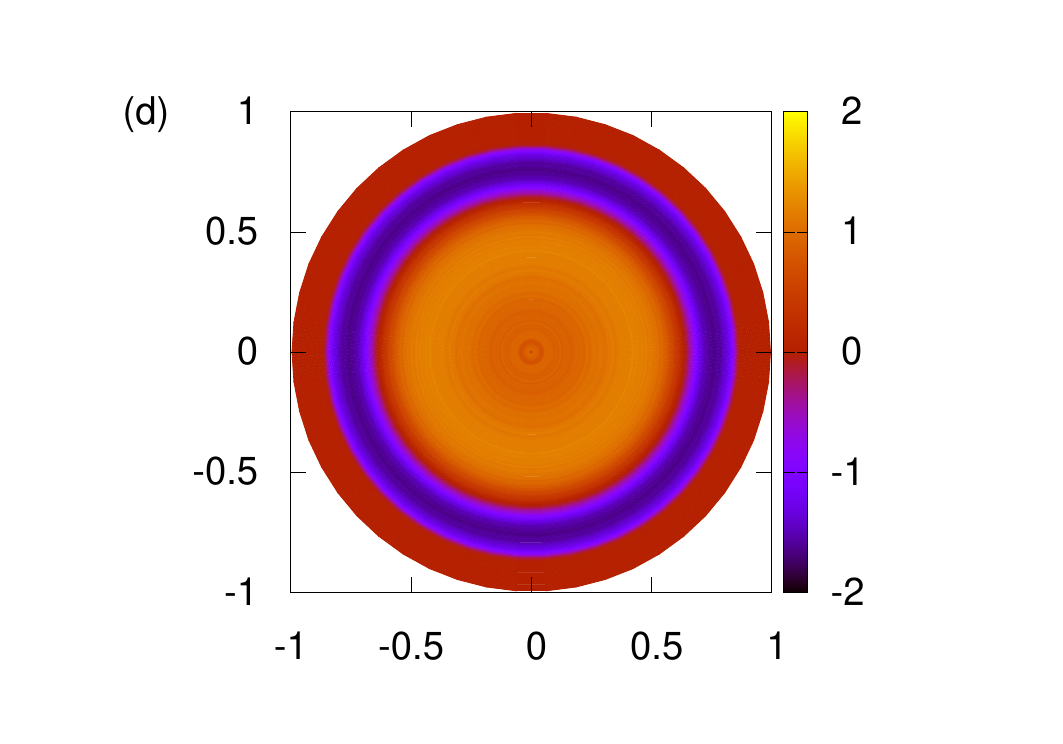}}
{\includegraphics[width=4.8cm]{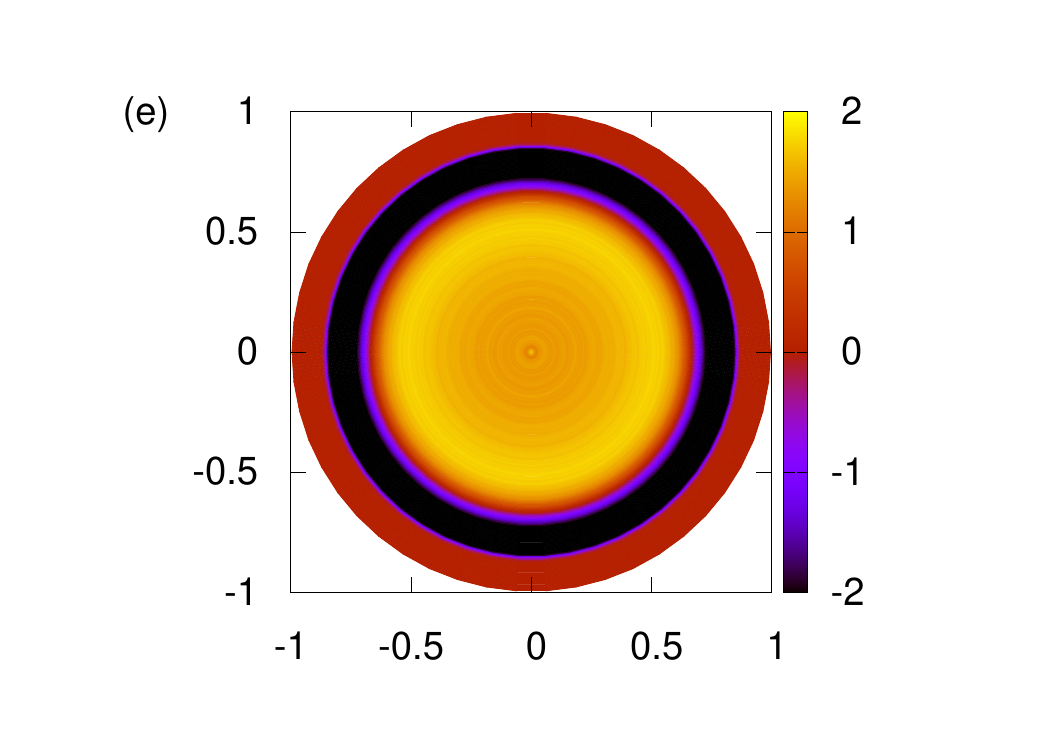}}
{\includegraphics[width=4.8cm]{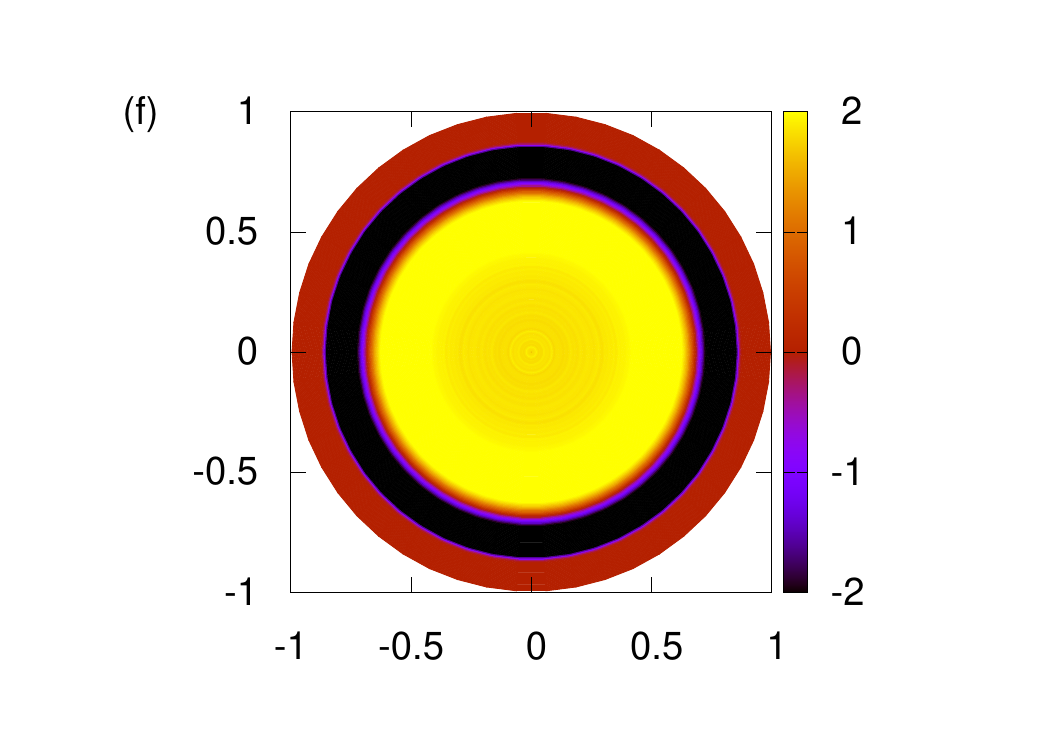}}
\caption{Color-map of the cross-section of the cylindrical confinement showing the net densities (normalized) for different values of the surface charge density: from 
(a)--(f) $\Sigma = 0, 0.04, 0.08, 0.12, 0.16, 0.20 ~C/m^2$. Values greater than zero on the color-map denote regions with excess positive charge (counterions) and values smaller 
than zero represent regions with excess negative charge (polyelectrolyte chains).}
\label{fig:heatmap_a}
\end{figure*}

Furthermore, as the surface charge density $\Sigma$ is increased, a re-arrangement of the polyelectrolyte and the counterion positions takes place inside the confinement, as 
depicted in the sequence of color-maps in Fig.\ref{fig:heatmap_a} showing the distribution of charges inside the confinement. For zero and low $\Sigma$, the excess counterion 
region is next to the inner surface of the confinement, whereas the excess polyelectrolyte region is next to the excess counterion region toward the interior of the confinement. 
This arrangement of charges reverses for larger $\Sigma$ values. The origin of this phenomenon, where there is an accumulation of ions near a charged surface of the same sign 
(positive in our case), is due to the interplay of short-range entropic effects and long-range Coulomb interactions, and is referred to as charge amplification (see 
\cite{bagchi2020surface} and references therein), since the presence of the positively charged counterions near the positively charged surface amplifies the charge on the 
confinement. Such an effect occurs only for the positively charged confinement and not for a negatively charged confinement \textcolor{blue}{(Fig. S5a, SI Appendix)}.

\subsection{Polyelectrolyte Separation and Polarization Effect}
Next, we look at changes in the mobility $\mu_P$, with and without surface polarization effect, when the number of charged monomers $N_m$ on the polyelectrolyte chains is varied. 
We take two polyelectrolyte solutions, one with $N_m = 40$ and the other with $N_m = 10$ charged monomers; both solutions have $N_c = 10$ polyelectrolyte chains. To study 
polarization effect, we consider that the confinement surface and the medium outside have a low dielectric constant $\epsilon_1 = 5$, whereas the solution inside the confinement 
has $\epsilon_2 = 80$. The polyelectrolyte mobility $\mu_P$, with polarization (w P) and without polarization (w/o P) effect, are shown in Fig. \ref{fig:poschconf}c. As can be 
seen, for the uncharged confinement ($\Sigma = 0$), both the solutions with $N_m = 10$ and $N_m = 40$, exhibit almost identical mobility. But when the confinement is charged 
($\Sigma > 0$), the mobilities of the two solutions can become very different. This indicates that a charged confinement can be employed to separate polyelectrolytes and other 
bio-molecules, based on the amount of charge they possess, by manipulating $\Sigma$. The higher mobility of the polyelectrolyte chains for $N_m = 10$ can be understood from the 
lower $f_{cs}$ in this case \textcolor{blue}{(Fig. S6, SI Appendix)}. This also justifies the more pronounced non-monotonicity of $\mu_P$ for $N_m = 10$ in Fig. 
\ref{fig:poschconf}c, because the polyelectrolyte chains get absorbed at a lower $\Sigma$ value, due to lower counterion screening.

To estimate the range of polyelectrolyte chain length over which one can potentially utilize this method for separation, we performed single chain simulations using chains of 
increasing $N_m$. The mobility data for $\Sigma = 0, 0.2$ and $0.5$ $C/m^{2}$ are shown in Fig. \ref{fig:poschconf}d. We vary the number of charged monomers between $10 \leq N_m 
\leq 1280$, and the mobility $\mu_P$ decreases steadily as the $N_m$ is increased for $\Sigma = 0.2$. This result suggests that separation of polyelectrolytes should be 
possible over a wide range of $N_m$ (over two orders of magnitude), and the separation process can be made efficient by tuning the SCD $\Sigma$ on the confinement. Note that, the 
monotonically decreasing mobility over the entire range of $N_m$ can be achieved only for an optimal $\Sigma$ value. As can be seen from Fig. \ref{fig:poschconf}d, for $\Sigma = 
0$, the change of $\mu_P$ is small and non-monotonic with $N_m$, whereas, for $\Sigma = 0.5$, $\mu_P$ is practically constant between $10 \leq N_m \lesssim 250$, and thus efficient 
electrophoretic separation over a wide range of $N_m$ is not possible for very low or very high $\Sigma$ values. Non-monotonic change in the polyelectrolyte mobility with the $N_m$ 
has been previously observed in simulations and experiments \cite{grass2008importance,hickey2013electrophoretic}, and it was attributed to long-range hydrodynamic interactions. In 
our case, there are no hydrodynamic interactions, and yet the mobility changes non-monotonically with polyelectrolyte length. 

For $\Sigma > 0$ in Fig. \ref{fig:poschconf}d, a simple power-law relation between $\mu_P$ and $N_m$ can be obtained from our simulation data to estimate the size dependence of the 
mobility, and is given by $\mu_P = a - bN_m^{\nu}$, where $a$, $b$, and $\nu$ are positive real numbers. The exponent $\nu$ is found to depend on the confinement charge density: 
for $\Sigma = 0.2$ and $0.5$, we obtain $(a, b, \nu) \cong (0.86, 1.6 \times 10^{-2}, 0.5)$ and $(0.61, 1.3 \times 10^{-5}, 1.41)$ respectively. For efficient polyelectrolyte 
separation with high resolution, one needs to optimize the values of both $b$ and $\nu$ by judiciously choosing the system parameters. Thus, without using sieving matrices or 
topological restrictions, a size dependent polyelectrolyte mobility can be obtained using a charged confinement which can be utilized to fractionate polyelectrolytes and other 
charged macromolecules. The size dependence originates from the differential friction imparted by the oppositely charged confinement surface on the polyelectrolyte chains. We have 
further verified that such a separation of polyelectrolytes is not possible for a confinement that has the same sign of charge as the polyelectrolyte chains \textcolor{blue}{(Fig. 
S7, SI Appendix)}.

However, the effect of surface polarization is clearly not appreciable in this model, as can be seen in Fig. \ref{fig:poschconf}c. The polyelectrolyte mobilities remain practically 
the same with and without polarization. In order to investigate if we can have substantial polarization effect that can be used for efficient polyelectrolyte separation, we study 
a slightly modified version of the present model, which we refer to as Model-II. The analysis of Model-II is described in the following sections.
\begin{figure}[htb]
\centering
{\includegraphics[width=7.000cm]{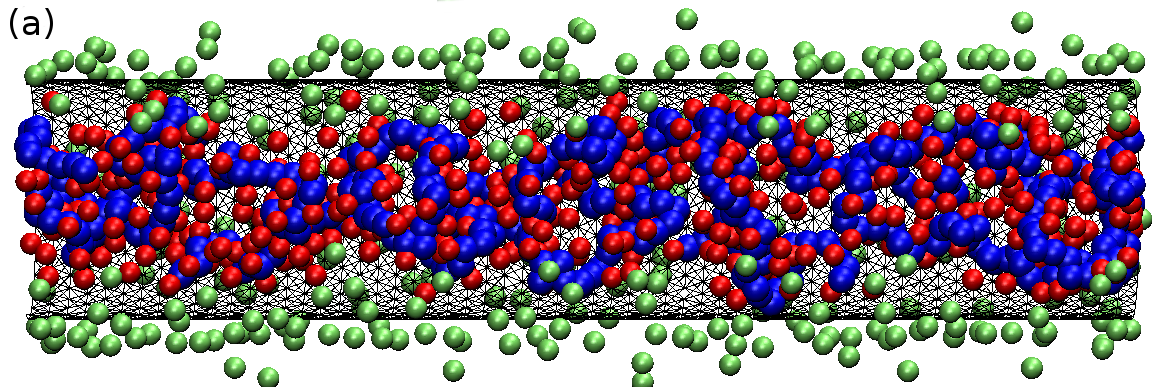}}
{\includegraphics[width=3.000cm]{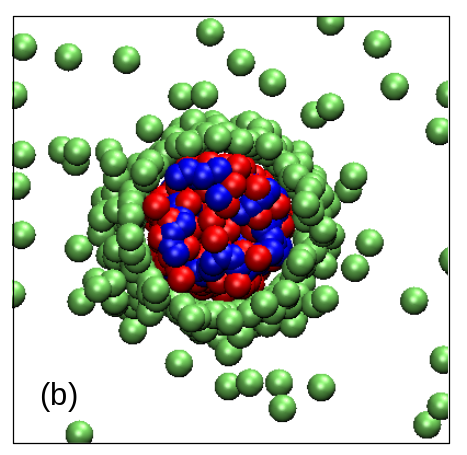}}
{\includegraphics[width=3.045cm]{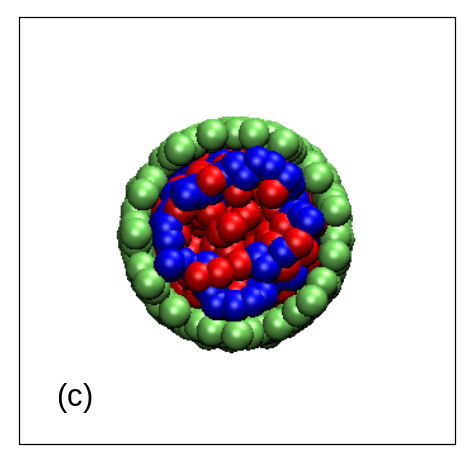}}
\caption{Model-II: (a) Same as Model-I (Fig. \ref{fig:model-I}), but with the surface counterions (green) outside the confinement. Typical configuration of the surface 
counterions (cross-sectional view): (b) without polarization and (c) with polarization, both for $\Sigma = 0.24$ $C/m^2$.}
\label{fig:model-II}
\end{figure}

\section{Simulation Results for Model-II}

In Model-II, we consider that all the surface counterions are {\it outside} the confinement, as shown in Fig. \ref{fig:model-II}. This slight modification imposes an additional 
constraint on the model, which is that the total charge inside the confinement is now zero, besides the overall charge neutrality of the entire system. As we demonstrate below, 
this additional constraint leads to intriguing consequences in the transport properties of the confined polyelectrolyte solution.

\subsection{Effect of surface polarization}
\begin{figure}[htb]
\hskip-0.1cm
{\includegraphics[width=5.90cm]{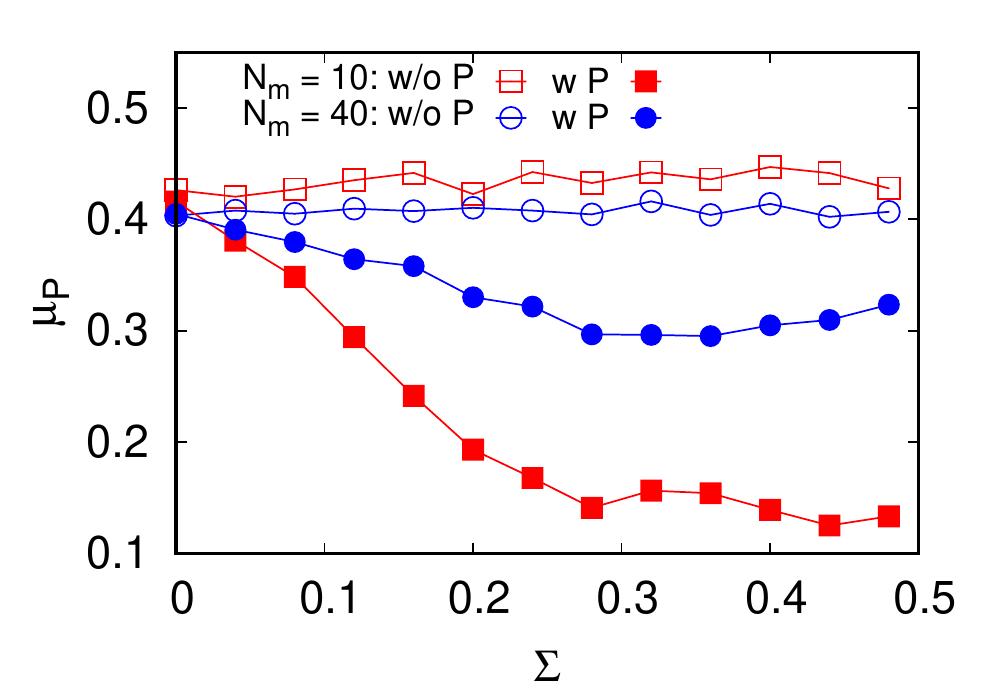}} \hskip-0.35cm
\caption{Polyelectrolyte mobility $\mu_P$ for two different solutions $N_m = 10$ and $N_m = 40$, without polarization (w/o P) and with polarization (w P). Here $N_c = 10$.}
\label{fig:sep_out}
\end{figure}

In order to investigate the role of surface polarization in Model-II, as before, we assume that the confinement and the medium outside the confinement have the same dielectric 
constant, $\epsilon_1 = 5$, whereas the polyelectrolyte solution has $\epsilon_2 = 80$. In Fig. \ref{fig:sep_out}, we show the mobility of the polyelectrolyte chains for two 
different solutions, one with $N_m = 10$ and the other with $N_m = 40$; we have $N_c = 10$ for both solutions. As anticipated, the effect of dielectric mismatch in this case is 
substantially larger compared to Model-I. Interestingly enough, the mobility $\mu_P$ is found to be independent of the SCD $\Sigma$ in the absence of polarization, but changes 
starkly when polarization effect is included.

\begin{figure}[htb]
\centering
{\includegraphics[width=5.90cm]{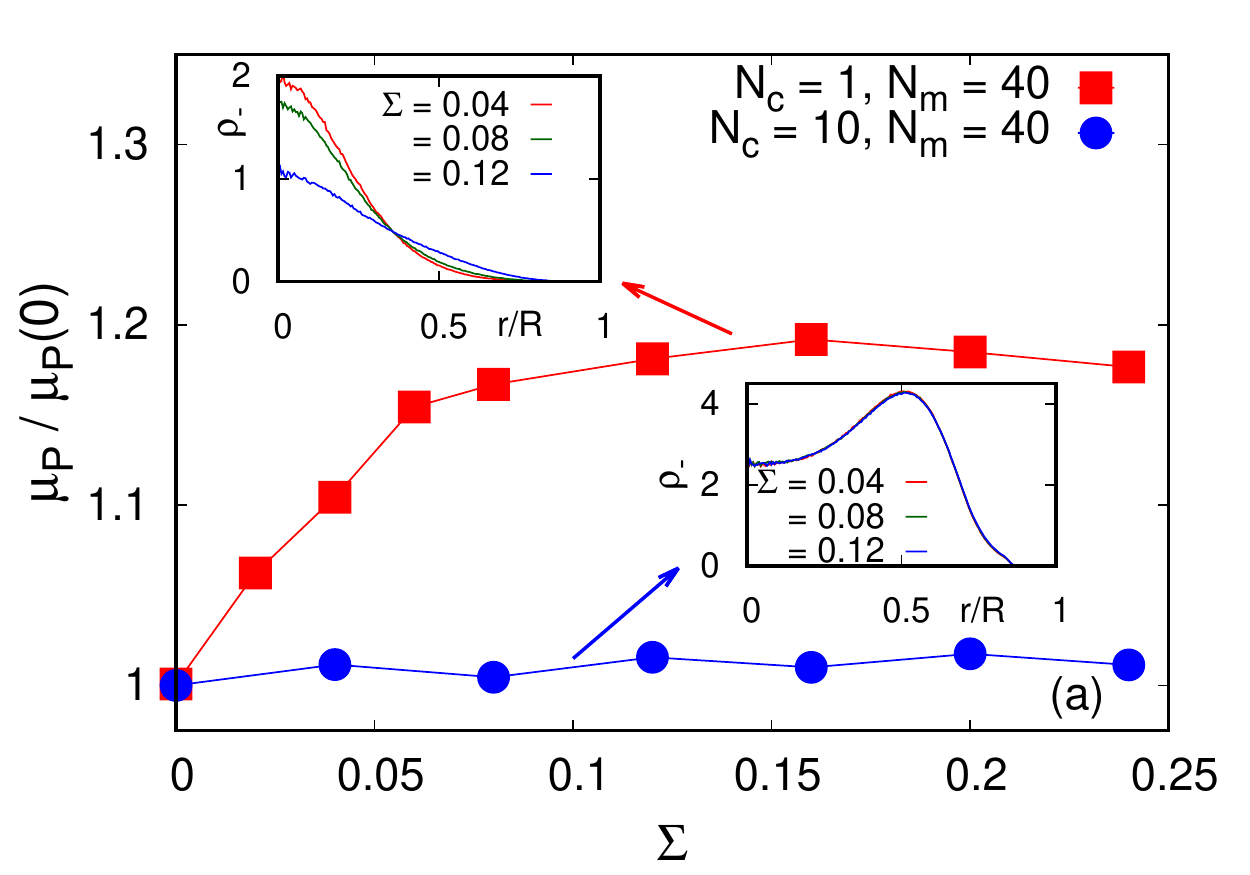}}\\
{\includegraphics[width=5.90cm]{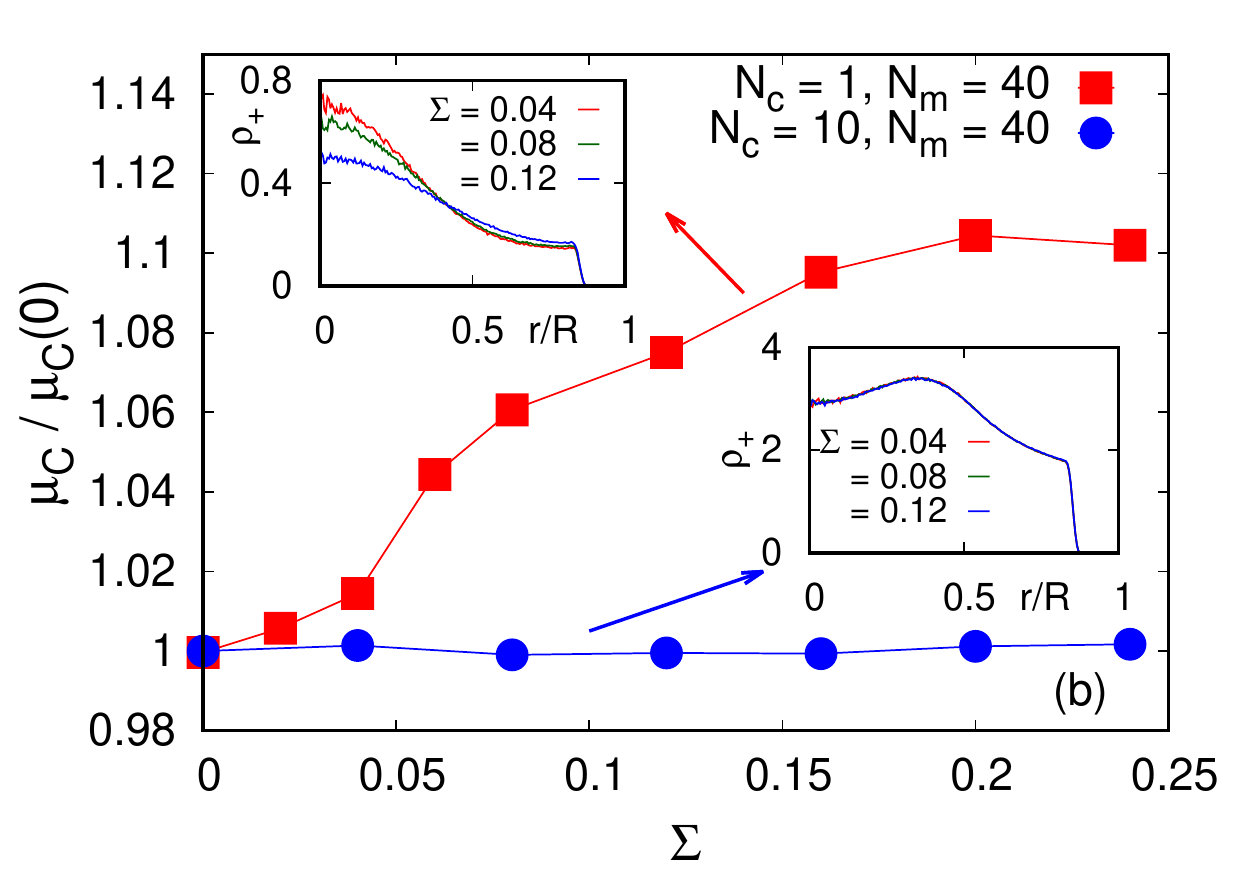}}
\caption{(a) Polyelectrolyte mobility and (b) counterion mobility for a single chain ($N_c = 1$) and multichain ($N_c = 10$) dynamics. In (a) and (b), the insets show the radial 
density of the monomers $\rho_-(r)$ and counterions $\rho_+(r)$ respectively. Here $N_m = 40$ for both the cases, and the mobilities are normalized by their $\Sigma = 0$ value.}
\label{fig:single}
\end{figure}

\begin{figure}[htb]
\hskip-0.25cm
{\includegraphics[width=4.50cm]{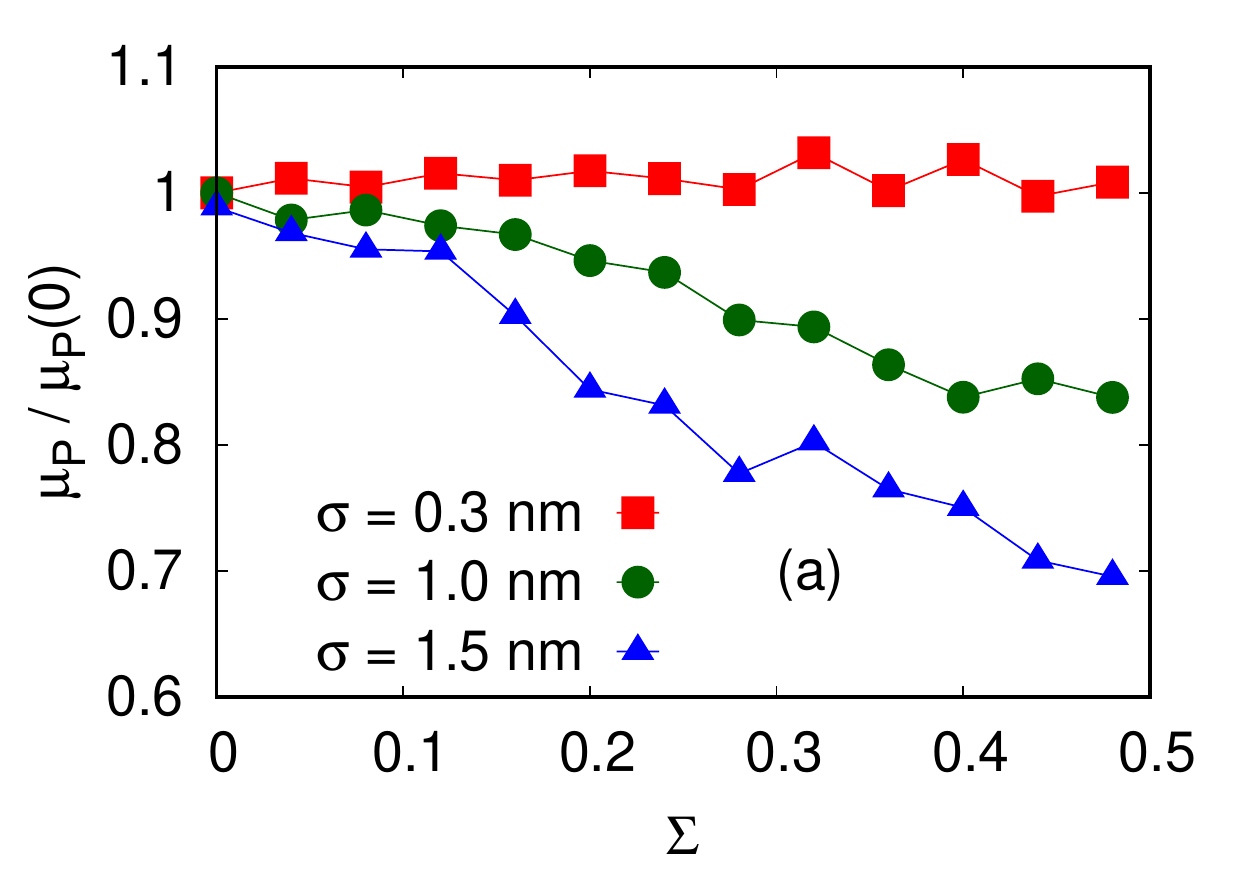}}\hskip-0.15cm
{\includegraphics[width=4.50cm]{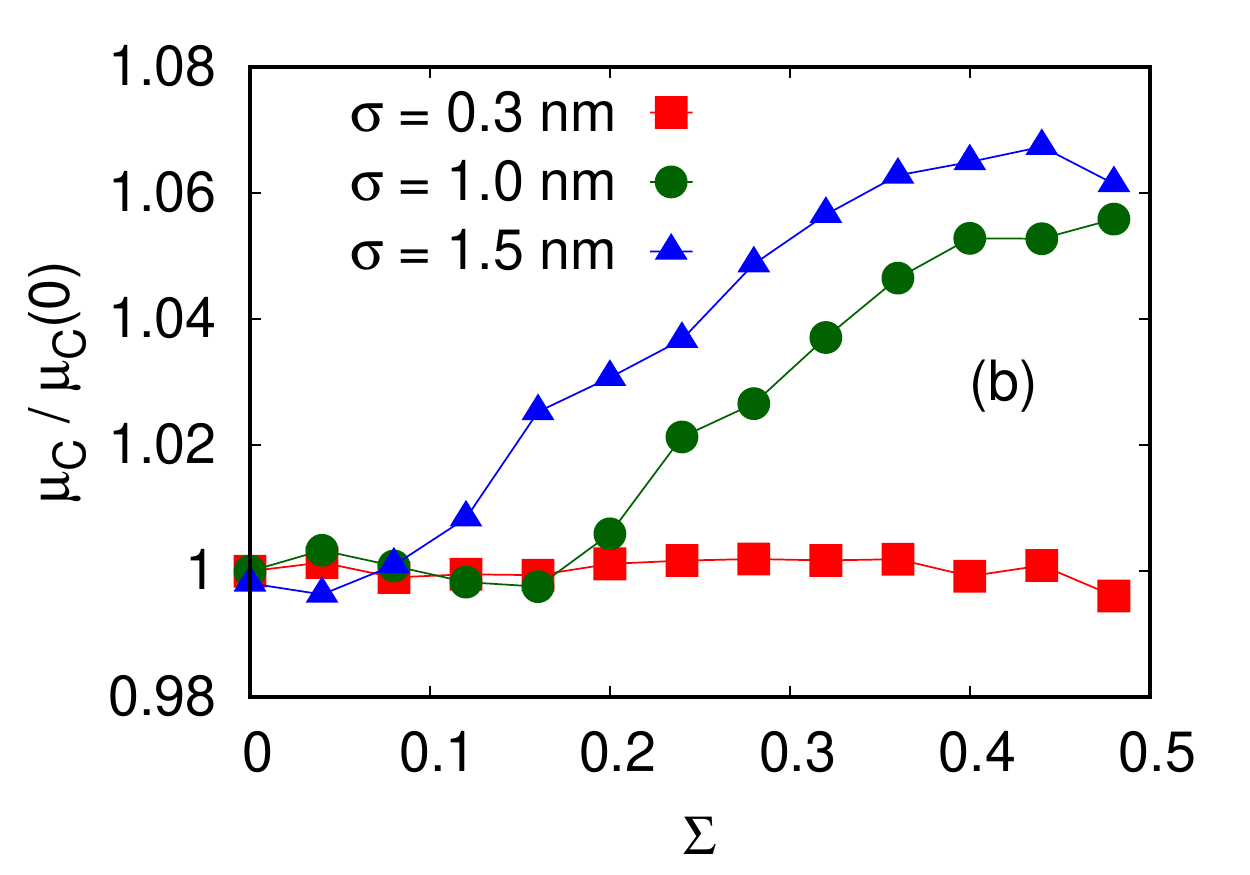}}
\caption{(a) Polyelectrolyte mobility and (b) counterion mobility for different bead size $\sigma = 0.3nm$, $1.0nm$, and $1.5nm$. The mobilities are normalized by their $\Sigma = 
0$ value. Here $N_m = 40$ and $N_c = 10$. The Bjerrum length is held constant at $l_B = 0.7nm$ for all the cases.}
\label{fig:beadsize}
\end{figure}

In order to explain the mobility trends in Fig. \ref{fig:sep_out}, a few comments are in order. Firstly, to understand why the mobilities are independent of the $\Sigma$, we note 
that the total charge inside the confinement is now zero, and by Gauss's law, the total electric flux $\phi = \int_S \vec E_s \cdot d \vec S$ inside a uniformly charged 
cylindrical confinement is zero, if there are no charges inside the confinement; $\vec E_s$ is the electric field generated by the charged surface. Thus, the positive and the 
negative charges inside the confinement cancel out and $\mu_P$ becomes independent of $\Sigma$. This independence of the mobilities on the surface charge density is also reflected 
in the net density $\rho_+ - \rho_-$ inside the confinement which is found to remain unaltered as $\Sigma$ is increased \textcolor{blue}{(Fig. S8, SI Appendix)}. Interestingly, we 
find that the mobilities $\mu_P$ and $\mu_C$ do depend on $\Sigma$ if there is a single polyelectrolyte chain ($N_c = 1$), along with its counterions, inside the confinement, as is 
shown in Fig. \ref{fig:single}. Thus a single polyelectrolyte chain and their counterions cannot produce the charge cancellation inside the confinement and the total electric flux 
$\phi$ inside the confinement is non-zero. This $\Sigma-$dependence can also be observed in the density profiles of the negatively charged monomers $\rho_-$ and the positively
charged counterions $\rho_+$ inside the confinement, for $N_c = 1$ and $N_c = 10$, as depicted in the insets of \ref{fig:single}a,b. Furthermore, we find that this cancellation of 
charges depends on the bead size used in the simulation, as can be seen in Fig. \ref{fig:beadsize}a,b that show polyelectrolyte and counterion mobilities for different bead size 
$\sigma$. For small bead size, the positive and negative charges cancel out, but for larger bead sizes, finite-size effect dominates electrostatic interactions, and charge 
cancellation is not achieved inside the confinement. Note that for all the curves in Fig. \ref{fig:beadsize}, the number of positive beads is exactly equal to the number of 
negative beads inside the confinement. These results also explain the observations of Ref. \cite{nguyen2019manipulation}, where the conformations of a single polyelectrolyte 
chain, for bead sizes larger than what we have used here, were found to strongly depend on the confinement surface charge density. Thus, a zero total charge inside the confinement 
is a {\it required} condition for the mobilities to be independent of $\Sigma$, but not a {\it sufficient} condition, since there are additional factors that decide whether or not 
the charges will cancel each other inside the confinement.

Secondly, one can explain the non-constant mobilities in the presence of polarization effect by noting that the sum total of all charges inside the confinement has to be zero in 
order to have a constant mobility. In the presence of a dielectric mismatch, the negative charges outside the confinement produce image charges inside the confinement, and 
therefore the total charge  $q_{total} = q_{real} + q_{image}$ is no longer zero inside the confinement in the presence of polarization. This leads to non-constant 
$\Sigma$-dependent mobilities when polarization effect is turned on. To further test that this argument is correct, we simulated a system with dielectric mismatch ($5|80$), but 
did not take into account the image charges (by turning off the ICC algorithm), and obtained a constant $\mu_P$ $versus$ $\Sigma$ (also for $\mu_C$ $versus$ $\Sigma$, 
\textcolor{blue}{Fig. S9a and b, SI Appendix}). This proves that the charges inside the confinement respond strongly to $\Sigma$ in the presence of polarization effect due 
to the image charges coming from the surface counterions outside the confinement.

Lastly, to explain the large effect of surface polarization in Model-II, we realize that the effect of surface polarization is very different between Model-I and Model-II. In 
Model-I, the effect of polarization is an additional repulsion of the charges away from the interface, which we refer to as the {\it dielectric confinement}. In Model-II, the 
surface counterions are in a medium of low dielectric constant ($\epsilon_1 = 5$) and hence they are strongly attracted toward the medium of high dielectric constant ($\epsilon_2 = 
80$) when polarization effect is taken into account. This can be seen from the representative simulation snapshots in Fig. \ref{fig:model-II}b and c, in the absence and presence 
of polarization effect respectively. This strong attraction of the negatively charged surface counterions and the confinement induces a large positive charge at the interface, 
particularly at larger values of $\Sigma$, for which there are a lot of counterions outside the confinement. This, in turn, leads to a strong attraction between the negatively 
charged polyelectrolyte chains and the positively charged confinement, whose bare charge density $\Sigma$ has been augmented by the additional positive induced charges. The 
polyelectrolyte chains release their counterions and get absorbed on the surface of the confinement (\textcolor{blue}{Figs. S9c and d, SI Appendix}), and hence, their mobility 
decrease in the presence of polarization effect, as can be observed in Fig. \ref{fig:sep_out}.

\subsection{Polyelectrolyte Separation} 
Next, in order to compare the separation of polyelectrolyte chains for Model-II, we perform simulations for two polyelectrolyte solutions with $N_m = 10$ and $N_m = 40$ as 
before, but this time with the surface counterions outside the confinement. The average mobility of the chains as $\Sigma$ is increased is shown in Fig. \ref{fig:sep_out}. Since 
the polyelectrolyte solution remains practically unaffected by $\Sigma$ in the absence of polarization effect, changing $N_m$ leads to a very small change in their mobility. As 
such, separation of polyelectrolytes is not achieved without polarization effect. However, the result changes drastically in the presence of polarization effect. As can be 
appreciated, there is a disparity in the mobilities of the two solutions, as $\Sigma$ is increased, and thus polyelectrolyte separation becomes possible when surface polarization 
effect is included.
\begin{figure}[htb]
\hskip-0.1cm
{\includegraphics[width=5.90cm]{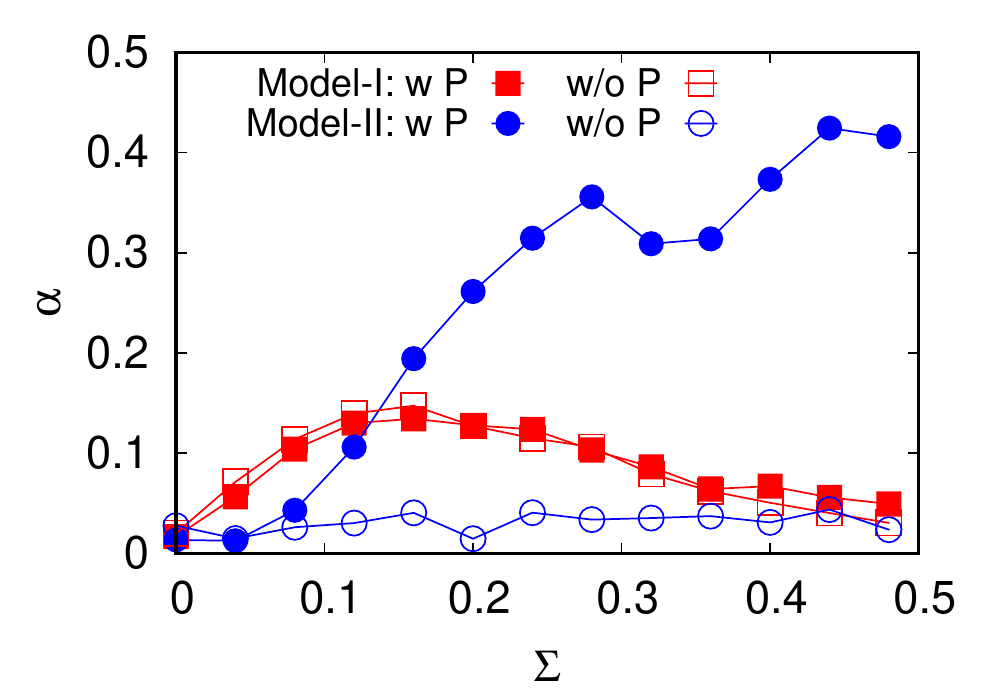}}
\caption{Separation efficiency $\alpha$ for Model-I and Model-II, with and without polarization effect. For Model-II with polarization effect (blue filled circles), one can 
achieve much higher $\alpha$ values.}
\label{fig:eff}
\end{figure}

To compare the efficiency of polyelectrolyte separation, achieved for the Model-I and Model-II, with and without polarization effect, we define an indicator for separation 
efficiency $\alpha = \dfrac{|\mu_{P,10} - \mu_{P,40}|}{|\mu_{P,10}| + |\mu_{P,40}|}$, where $\mu_{P,10}$ ($\mu_{P,40}$) is the chain mobility for the polyelectrolyte solution with 
$N_m = 10$ ($N_m = 40$) charged monomers. This is shown in Fig. \ref{fig:eff}. Clearly, in the presence of polarization effect, one can achieve much larger separation 
efficiency for Model-II as compared to Model-I.

\begin{figure}[htb]
\hskip-0.4cm
{\includegraphics[width=4.6cm]{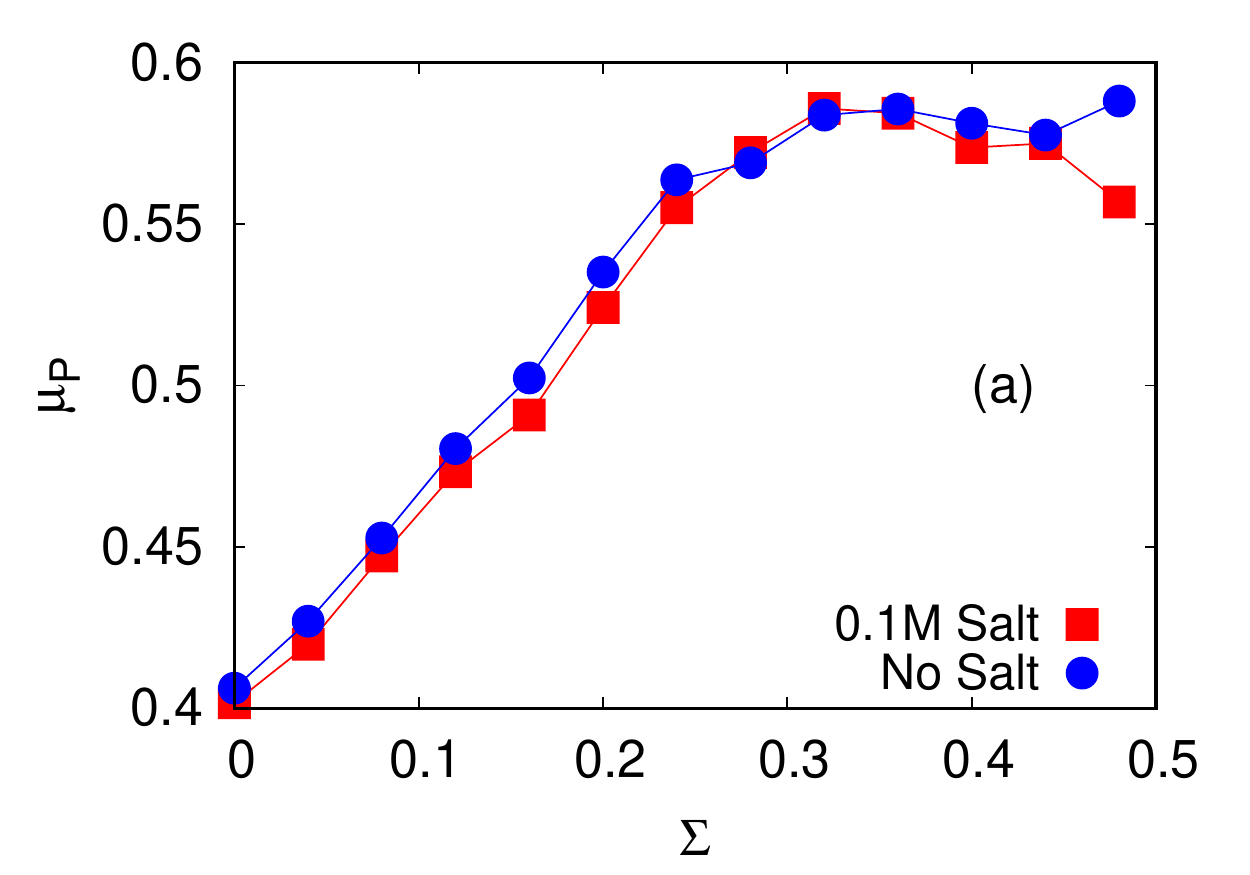}}\hskip-0.25cm
{\includegraphics[width=4.6cm]{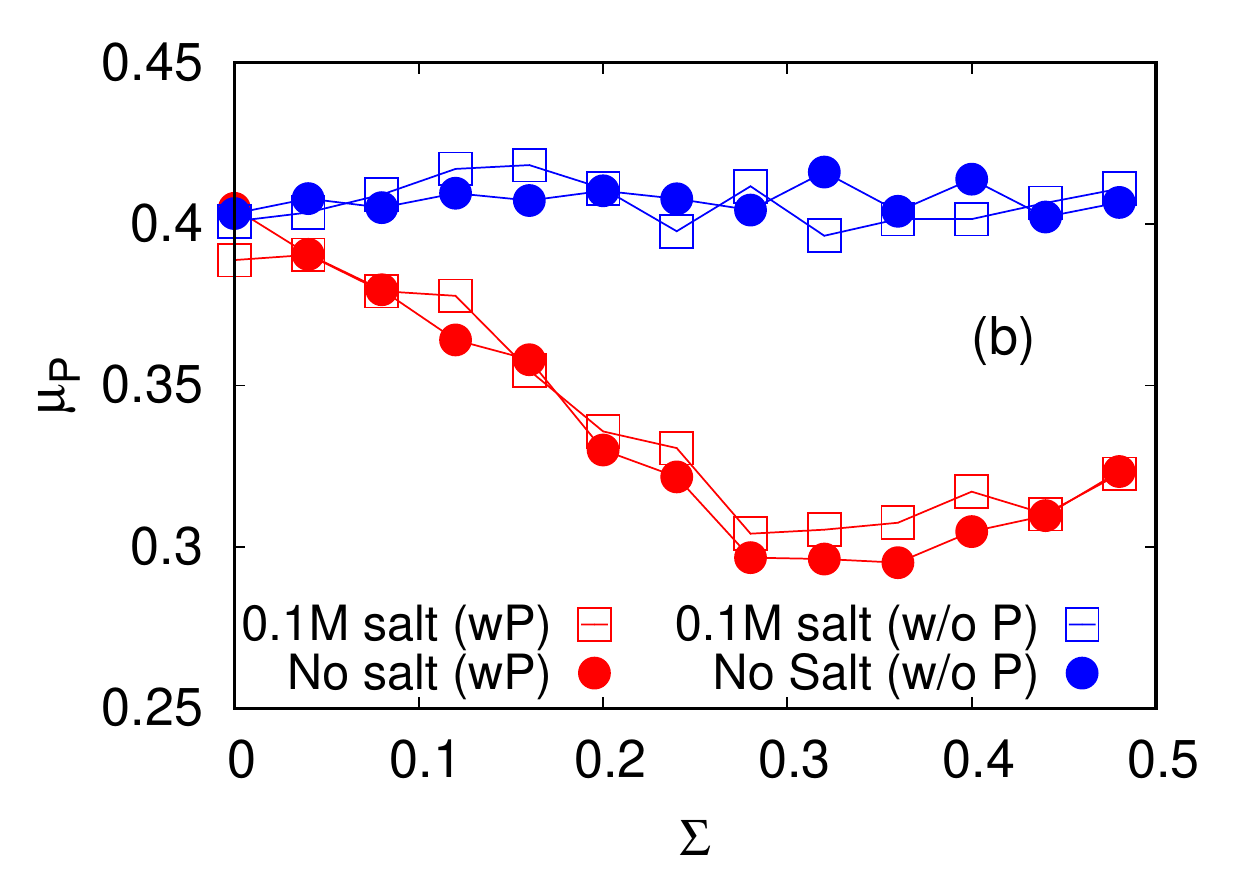}}
\caption{Polyelectrolyte mobility in the presence of $0.1M$ monovalent salt for (a) Model-I and (b) Model-II. For Model-I, we show data only for the case without polarization, and 
for Model-II, we show data both with polarization (wP) and without polarization (w/o P) effect. For these simulations $N_m = 40$ and $N_c = 10$.}
\label{fig:salt}
\end{figure}

We also performed some simulations in the presence of monovalent salts for both the models. The results for $0.1M$ salt are shown in Fig. \ref{fig:salt} along with the data in 
the absence of salt. For Model-I, we present data only for the case without polarization, since the effect of polarization is negligible in this case (Fig. \ref{fig:poschconf}c). 
For Model-II, the effect of polarization is dramatically different but the results remain the same with and without the addition of salt. Thus, for both the models, we find that 
the transport properties are robust and adding salt seems to have no visible effect.

\section{Conclusion} 

In conclusion, the transport properties of confined polyelectrolyte solution can be conveniently modulated by tuning the surface charge density on the confinement. For Model-I, 
which has the surface counterions dissolved in the confined polyelectrolyte solution, we have shown that the mobilities of the polyelectrolytes and their counterions vary 
non-monotonically with the surface charge density on the confinement. One can achieve separation of polyelectrolytes over a wide range of charge (size) by exploiting the 
electrostatic friction from the charged confinement surface, but the effect of surface polarization is found to be negligibly small in this case.
For Model-II, which has the surface counterions in the low dielectric media, we find that the polyelectrolyte and counterion mobilities are independent of the surface charge 
density $\Sigma$ in the absence of polarization effect, but become $\Sigma-$dependent in the presence of polarization. In this case, the effect of polarization is significantly 
higher compared to Model-I due to the presence of charges in a medium of low dielectric constant. We have further shown that by exploiting the effect of surface polarization due to 
mismatch of dielectric constants, enhanced separation of polyelectrolytes can be achieved in this case.
The polyelectrolyte transport properties are found to be quite robust in the presence of moderately high concentrations of monovalent salt inside the confinement.
Thus, without requiring a sieving medium, one can fractionate polyelectrolytes, such as DNA, using the differential friction between polyelectrolytes and an oppositely charged 
confinement by tuning its surface charge density. The charged nanochannel studied in this work should be possible to realize by manipulating standard materials that are frequently 
used in nanotechnology, such as polydimethylsiloxane (PDMS), which ionize and acquire surface charge when in contact with a suitable solvent. Our molecular simulations reveal the 
rich underlying physics of a driven confined polyelectrolyte solution and the results presented here will be relevant for designing efficient DNA sequencers and other nanofluidic 
devices.

\begin{acknowledgments}
This work was supported by the Sherman Fairchild Foundation. We also thank QUEST at Northwestern University for providing computational facilities.
\end{acknowledgments}

\section*{Data availability}
The data that support the findings of this study are available from the corresponding author upon reasonable request.


\end{document}